\documentclass[options]{JHEP3}
\title{Relating the Green-Schwarz and Pure
Spinor Formalisms for the Superstring}
\author{Nathan Berkovits and D\'{a}fni Z. Marchioro \\
Instituto de
F\'{\i}sica Te\'{o}rica, Universidade Estadual
Paulista\\ Rua Pamplona 145,
01405-900, S\~{a}o Paulo, SP, Brasil \\E-mail:
\email{nberkovi@ift.unesp.br,
dafni@ift.unesp.br}}

\abstract{Although it is not known how to covariantly
quantize
the Green-Schwarz (GS) superstring, there exists a
semi-light-cone gauge
choice in which the GS superstring can be quantized in
a conformally invariant manner. In this paper, we
prove that BRST
quantization of the GS superstring in semi-light-cone
gauge is
equivalent to BRST quantization using the pure spinor
formalism for
the superstring}


\preprint{IFT-P.058/2004}

\begin{document}

\section{Introduction}

Four years ago, a new manifestly super-Poincar\'e
covariant formalism
was introduced for quantizing the superstring \cite{superstring}.
This formalism has been recently used for computing
covariant
multiloop amplitudes \cite{multiloop} and for
quantization in an $AdS_5\times S^5$ Ramond-Ramond
background \cite{ads}.
The main new ingredient in this formalism is a BRST
operator constructed
from fermionic Green-Schwarz (GS) constraints and a
pure spinor bosonic
ghost.

At the present time, a geometrical understanding of
this ``pure spinor''
BRST operator is still lacking.\footnote{However,
there are some indications
that the pure spinor BRST operator can be interpreted
as a twisted N=2
worldsheet supersymmetry generator using either a
``twistor-superstring''
formalism \cite{tonin} or a WZNW model \cite{policastro}.} It is
therefore important
to try to relate
the pure spinor BRST operator with BRST operators
which appear in
other formalisms for the superstring. In reference \cite{rns},
this was done for the
RNS formalism where the pure spinor BRST operator was
related to the
sum of the RNS BRST operator and $\eta$ ghost. It was
also shown in \cite{cohomology}
that the cohomology of the pure spinor BRST operator
reproduces
the light-cone GS spectrum.\footnote{There also exist
``extended''
versions of the pure spinor formalism in which the
pure spinor constraint
on the bosonic ghost is relaxed
\cite{grassi}\cite{kazama}.
The BRST operator in these extended pure spinor
formalisms have
been related to the RNS BRST operator \cite{aisaka}, to the
light-cone GS
spectrum \cite{kaza}, and to the original pure spinor BRST operator
\cite{grassitwo}.}

Although it is not known how to covariantly quantize
the GS superstring,
one can choose a semi-light-cone gauge in which
$\kappa$-symmetry is
fixed but conformal invariance is preserved \cite{carlip-kallosh}. The
worldsheet
action is quadratic in semi-light-cone gauge,
so quantization is straightforward
using the BRST method.
Although Lorentz invariance is not manifest in this
gauge, one can
construct Lorentz generators whose algebra closes up
to
a BRST-trivial quantity.

In this paper, the pure spinor
BRST operator will be
related to the semi-light-cone GS BRST operator by
a similarity
transformation, proving the equivalence of the
cohomologies.
Note
that a similar result was obtained earlier for the
$d=10$ \cite{ictp} and $d=11$
superparticle \cite{anguelova} in semi-light-cone
gauge. However, relating the BRST operators
for the superstring
is more complicated than for the superparticle
because of normal-ordering subtleties. It would
be interesting to try to further generalize this
equivalence proof to the
$d=11$ supermembrane for which pure spinor \cite{membrana} and
semi-light-cone gauge
descriptions exist at least at the classical level.

In section 2 of this paper, we review the equivalence
proof
for the $d=10$
superparticle.
And in section 3, we generalize this equivalence proof
to the superstring.

\section{Review of Equivalence Proof for Superparticle}

\subsection{Brink-Schwarz superparticle in the
semi-light-cone gauge}

The $d=10$ Brink-Schwarz superparticle is described by
the
action

\begin{equation}
S = \int d\tau \left(\dot{x}^{m}P_{m}
-\frac{i}{2}(\dot{\theta}\gamma^{m}\theta)P_{m} +
eP^{m}P_{m}\right) \:,
\label{superparticle}
\end{equation}

\noindent where $P_{m}$ is the conjugate momentum for
$x^{m}$, $\theta^{\alpha}$ is
a spinor of $SO(9,1)$, $e$ is a Lagrangian multiplier
which enforces the
mass-shell condition, $\gamma^{m}_{\alpha\beta}$
are $16\times16$
symmetric gamma matrices which satisfy
$\gamma^{(m}_{\alpha \beta}\gamma^{n)\beta \lambda} =
2\eta^{mn}\delta^{\lambda}_{\alpha}$, and we use the
metric convention $\eta^{mn} =
\mbox{diag}(-1,1,1,\ldots)$.

The action (\ref{superparticle}) is spacetime
supersymmetric and is also
invariant under local kappa transformations, which are
generated by the
first-class part of the fermionic
constraints

\begin{equation}
d_{\alpha} = p_{\alpha} +
2P^{m}(\gamma_{m}\theta)_{\alpha} \:,
\label{constraint}
\end{equation}

\noindent which satisfy the Poisson brackets

\begin{equation}
\{d_{\alpha}, d_{\beta} \} =
4P_{m}\gamma^{m}_{\alpha\beta} \:.
\label{primeira-classe}
\end{equation}

\noindent Eight of the $d_\alpha$ constraints
are first-class
and the other eight are second-class.
There is no simple
way of covariantly
out the first-class constraints, preventing covariant
quantization.
Nevertheless, it is possible
to use a non-Lorentz covariant gauge fixing to
quantize this theory
by imposing the
condition
$(\gamma^{+}\theta)_{\alpha} = 0$ with the assumption
that
$P^+\neq 0$ (where $\gamma^{\pm}
= \gamma^{0} \pm
\gamma^{9}$ and $P^{\pm}
= P^{0} \pm P^{9}$). In this
semi-light-cone gauge,
the action takes the form

\begin{equation}
S = \int d\tau \left(\dot{x}^{m}P_{m} +
\frac{i}{2}\dot{S}_{a}S_{a} +
eP^{m}P_{m}\right) \:,
\label{semilight}
\end{equation}

\noindent where $S_{a} =
\sqrt{\frac{P^{+}}{2}}(\gamma^{-}\theta)_{a}$ and
$a=1$ to 8 is an $SO(8)$ chiral spinor index.
Canonical quantization of
(\ref{semilight}) implies that $\{S_{a}, S_{b}\} =
\delta_{ab}$ and therefore
$\sqrt{2} S_a$ acts like a spinor version of $SO(8)$
Pauli
matrices that satisfy

\begin{equation}
\sigma^{j}_{a\dot{a}}\sigma^{j}_{b\dot{b}} +
\sigma^{j}_{b\dot{a}}\sigma^{j}_{a\dot{b}} =
2 \delta_{ab}\delta_{\dot{a}\dot{b}} \:.
\end{equation}

Making use of the usual BRST method, the action
(\ref{semilight}) can be
gauge-fixed to

\begin{equation}
S = \int d\tau \left(\dot{x}^{m}P_{m} +
\frac{i}{2}\dot{S}_{a}S_{a}
-\frac{1}{2}P^{m}P_{m} + i\dot{c}b\right)  \:,
\label{particle}
\end{equation}

\noindent where the BRST charge is

\begin{equation}
Q = cP^{m}P_{m} \:.
\end{equation}

\noindent Note that $\kappa$-symmetry
ghosts do not propagate in semi-light-cone gauge,
so the BRST operator only involves the $(b,c)$
reparametrization ghosts.

Because semi-light-cone gauge is not manifestly
Lorentz invariant,
Lorentz transformation which change the gauge-fixing
condition
$(\gamma^+\theta)_\alpha=0$ need
to include a compensating $\kappa$-transformation.
The resulting Lorentz generators are

\begin{eqnarray}
N^{+-} &=& -ix^{+}P^{-} +ix^{-}P^{+}\:, \nonumber\\
N^{i+} &=& -ix^{i}P^{+} +ix^{+}P^{i} \:, \nonumber\\
N^{i-} &=& -ix^{i}P^{-} +ix^{-}P^{i}
-\frac{(S\sigma^{i})_{\dot{a}}(S\sigma^{j})_{\dot{a}}P^{j}}{2P^{+}}
\:,
\nonumber\\
N^{ij} &=& -ix^{i}P^{j} +ix^{j}P^{i} -
\frac{1}{4}(S_a\sigma^{ij}_{ab}S_b) \:,
\label{lorentz}
\end{eqnarray}

\noindent where $x^{\pm}$, $P^{\pm}$ are light-cone
variables, and $i,j = 1$
to 8. Using the canonical commutation relations
$\{S_a,S_b\}=\delta_{ab}$
and $[P^m,x^n]= -i\eta^{mn}$, one finds that the
Lorentz generators of (\ref{lorentz})
satisfy the usual $SO(9,1)$
Lorentz algebra except for $[N^{i-},N^{j-}]$, which
satisfies

\begin{equation}
[N^{i-},N^{j-}] = \left[{Q},
-\frac{b(S\sigma^{i})_{\dot{a}}(S\sigma^{j})_{\dot{a}}}{(P^{+})^2}\right]\:.
\end{equation}

Since $[N^{i-},N^{j-}]$ is BRST-trivial, the Lorentz
algebra closes
up to a gauge transformation when acting on
BRST-invariant states.
In other words, $QV=0$ implies that
$[N^{i-},N^{j-}] V = Q\Omega$ for some $\Omega$.

\subsection{Equivalence between Brink-Schwarz
superparticle and pure spinor
superparticle}

In this subsection, it will be shown that the action and
BRST operator
for the semi-light-cone Brink-Schwarz superparticle
in the previous subsection
are related
to the action
\begin{equation}
S = \int d\tau \left(\dot{x}^{m}P_{m}
-\frac{1}{2}P^{m}P_{m} +
i\dot{\theta}^{\alpha} p_{\alpha} +i\dot\lambda^\alpha
w_\alpha\right)\:,
\label{pure}
\end{equation}
and BRST operator
\begin{equation}
Q = \lambda^{\alpha}d_{\alpha} \:
\label{charge}
\end{equation}
for the pure spinor version of the superparticle \cite{particula} where
$(\lambda^\alpha, w_\alpha)$ are bosonic ghosts
satisfying the
pure spinor constraint
\begin{equation}
\lambda^{\alpha}\gamma^{m}_{\alpha\beta}\lambda^{\beta}
= 0 \:.
\label{cond}
\end{equation}

To relate the actions of (\ref{particle}) and
(\ref{pure}), we shall first introduce a
new pair
of fermionic variables
$(\theta^{\alpha},p_{\alpha})$ which
are not related with
$S_a$. And to prevent physical states from depending
on these new
variables, we shall also introduce a new
gauge invariance which allows $\theta^\alpha$ to be
gauged to zero.
This new gauge invariance will be generated by the
first-class constraints
\begin{equation}
\widehat{d}_{\alpha} = d_{\alpha} +
\frac{(\gamma_{m}\gamma^{+}S)_{\alpha}P^{m}}{\sqrt{P^{+}}}
\:,
\label{hat}
\end{equation}
where $d_{\alpha} = p_{\alpha}+
2P^{m}(\gamma_{m}\theta)_{\alpha}$.
Or using $SO(8)$ notation,

\begin{eqnarray}
\widehat{d}_{a} &=& d_{a} + 2S_{a} \sqrt{P^+}
\:,\nonumber\\
\widehat{{d}}_{\dot{a}} &=& {d}_{\dot{a}} -
\frac{2 P^{i}(S\sigma^{i})_{\dot{a}}}{\sqrt{P^{+}}}\:,
\label{hat-gsparticle}
\end{eqnarray}

\noindent where $a, \dot{a} = 1$ to 8 are $SO(8)$ chiral
and antichiral spinor indices.

Using (\ref{primeira-classe})
and the anticommutation relation of the $S_a$'s, one
can check that the
constraints
satisfy the first-class algebra
\begin{equation}
\{\widehat{d}_{\alpha},\widehat{d}_{\beta}\} =
\frac{2P^{m}P_{m}\gamma^{+}_{\alpha\beta}}{P^{+}} \:,
\end{equation}
or in $SO(8)$ notation,
\begin{equation}
\{\widehat{d}_{a},\widehat{d}_{b}\} =
\{\widehat{d}_{a},\widehat{ d}_{\dot b}\} = 0,\quad
\{\widehat{ d}_{\dot a},\widehat{ d}_{\dot b}\} =
\frac{4P^{m}P_{m}\delta_{\dot a \dot b}}{P^{+}} \:.
\end{equation}

\noindent So the semi-light-cone superparticle action
which includes
the new variables and new gauge invariances is

\begin{equation}
S = \int d\tau \left(\dot{x}^{m}P_{m} + eP^{m}P_{m} +
i\dot{\theta}^{\alpha} p_{\alpha} +
\frac{i}{2}\dot{S}_{a}S_{a} +
f^{\alpha}\widehat{d}_{\alpha}\right) \:,
\end{equation}

\noindent where $f^{\alpha}$ are fermionic Lagrange
multipliers related to
the constraint (\ref{hat}). To return to the original
action in the
semi-light-cone gauge, $\widehat{d}_{\alpha}$ can be used
to gauge
$\theta^{\alpha}=0$, recovering (\ref{semilight}).
Using the usual BRST
method, one can gauge fix the action above and obtain

\begin{equation}
S = \int d\tau \left(\dot{x}^{m}P_{m}
-\frac{1}{2}P^{m}P_{m} +
i\dot{\theta}^{\alpha} p_{\alpha} +
\frac{i}{2}\dot{S}_{a}S_{a} +
i\dot{\widehat{\lambda}}^{\alpha}\widehat{w}_{\alpha} +
i\dot{c}b\right) \:,
\end{equation}

\noindent together with the BRST charge

\begin{equation}
\widehat{Q} = \widehat{\lambda}_{a} \widehat d_{a} +
\widehat{\lambda}_{\dot{a}}\widehat{ d}_{\dot{a}}
+ c\left(-4P^{-} +
\frac{4P^{i}P^{i}}{P^{+}}\right) -
\frac{1}{2}\widehat{\lambda}_{\dot{a}}\widehat{\lambda}_{\dot{a}}b
\label{carga1}
\end{equation}

\noindent where $\widehat{\lambda}_{\alpha}$ is an
unconstrained bosonic
spinor ghost related to
the gauge fixing $f^{\alpha}=0$, $\widehat{w}_{\alpha}$ is
its conjugate momentum,
and $a, \dot{a},i=1$ to 8.
Note that
one could rescale the $(b,c)$ ghosts
by factors of $P^+$ as
$$b\to {b\over {P^+}}, \quad c\to c P^+,$$
so that the $c$ ghost would multiply $P_m P^m$ in
the BRST operator. But it will be more convenient for
generalization
to the superstring
to leave the superparticle BRST operator in the form
of (\ref{carga1}).
This means that $b$ and $c$ are not invariant under
Lorentz transformations
which change the value of $P^+$.

As before, Lorentz invariance is not manifest but one
can define
Lorentz generators which commute with the BRST
operator
and whose algebra closes up to a BRST-trivial
quantity.
The explicit expression for the Lorentz generators is

\begin{eqnarray}
N^{+-} &=& -ix^{+}P^{-} + ix^{-}P^{+} +\theta_{a}p_{a}
- {\theta}_{\dot{a}}{p}_{\dot{a}} +
\widehat{\lambda}_{a}\widehat{w}_{a} -
\widehat{\lambda}_{\dot{a}}\widehat{ w}_{\dot{a}} - 2bc
\:,\nonumber\\
N^{i+} &=& -ix^{i}P^{+} + ix^{+}P^{i} -
(p\sigma^{i}{\theta}) +
(\widehat{w}\sigma^{i}\widehat{\lambda}) \:,\nonumber\\
N^{ij} &=& -ix^{i}P^{j} +ix^{j}P^{i} +
\frac{1}{2}(\theta\sigma^{ij}p) +
\frac{1}{2}({\theta}{\overline\sigma}^{ij}{p}) +
\frac{1}{2}(\widehat{\lambda}\sigma^{ij}\widehat{w}) +
\frac{1}{2}(\widehat{\lambda}\overline{\sigma}^{ij}
\widehat{ w})
-
\frac{1}{4}(S\sigma^{ij}S) \:, \nonumber\\
N^{i-} &=& -ix^{i}P^{-} + ix^{-}P^{i} +
\frac{b(S\sigma^{i}\widehat{\lambda})}{2\sqrt{P^{+}}} -
\frac{(S\sigma^{i})_{\dot{a}}(S\sigma^{j})_{\dot{a}}P^{j}}{2P^{+}}
\nonumber\\
&+& (\theta\sigma^{i}{p}) + (\widehat{\lambda}
\sigma^{i}\widehat{ w}) -
\frac{2bcP^{i}}{P^{+}} \:,
\end{eqnarray}
\noindent where
$$(A \sigma^i B) = A_a \sigma^i_{a \dot b} B_{\dot b},\quad
(A \sigma^{ij} B) =\frac{1}{2} A_a
\sigma^{[i}_{a  \dot c}\sigma^{j]}_{b \dot c} B_b ,\quad
(A \overline\sigma^{ij} B) =\frac{1}{2} A_{\dot a}
\sigma^{[i}_{c  \dot a}\sigma^{j]}_{c \dot b} B_{\dot b} .$$

\noindent The generators obey the usual Lorentz
algebra, except for the
commutator of $N^{i-}$ with $N^{j-}$ which satisfies

\begin{equation}
[N^{i-},N^{j-}] = \left[\widehat{Q}, -
\frac{b(S\sigma^{i})_{\dot{a}}(S\sigma^{j})_{\dot{a}}}{P^{+}}\right]\:,
\end{equation}

\noindent indicating that the algebra closes up to a
gauge transformation on on-shell states.

It will now be shown that the BRST operator of
(\ref{carga1}) is related by a similarity
transformation to the pure
spinor BRST operator $Q=\lambda^\alpha d_\alpha $
\cite{ictp}.
The first step will be to show that the cohomology of
the BRST operator in (\ref{carga1}) is
equivalent to the cohomology of
$Q'=\widehat{\lambda}_a\widehat{d}_a + {\lambda}_{\dot a} \widehat
{ d}_{\dot a}$ in
a Hilbert space without the $(b,c)$ ghosts and with
the condition ${\lambda}_{\dot a}{\lambda}_{\dot a}=0$.

Suppose that
a state $V$ is in the cohomology of $Q'$. Then $V$ is annihilated
by the operator
\begin{equation}
{Q''} = \widehat{\lambda}_{a} \widehat d_{a}
+\widehat{\lambda}_{\dot{a}}\widehat{ d}_{\dot{a}}
\label{carga5}
\end{equation}

\noindent up to terms proportional to
$\widehat{\lambda}_{\dot{a}}\widehat{\lambda}_{\dot{a}}$,
i.e.,

\begin{equation}
Q^{''}V =
\widehat{\lambda}_{\dot{a}}\widehat{\lambda}_{\dot{a}}W \:,
\label{abovee}
\end{equation}

\noindent for some $W$. Since

\begin{equation}
(Q^{''})^{2} =
\frac{2\widehat{\lambda}_{\dot{a}}\widehat{\lambda}_{\dot{a}}P^{m}P_{m}}{P^{+}}
\:,
\end{equation}

\noindent (\ref{abovee}) implies that

\begin{equation}
Q^{''}W = \frac{2P^{m}P_{m}V}{P^{+}} \:.
\end{equation}

\noindent which implies that
the state
$\widehat{V}=V + 2cW$ is annihilated by $\widehat{Q}$.
And if $V$ is
BRST-trivial up to terms proportional to
$\widehat{\lambda}_{\dot{a}}\widehat{\lambda}_{\dot{a}}$,
i.e.,

\begin{equation}
V = Q^{''}\Omega +
\widehat{\lambda}_{\dot{a}}\widehat{\lambda}_{\dot{a}}Y \:
\end{equation}

\noindent for some $\Omega$ and $Y$, then $\widehat{V} = V
+ 2cW =
\widehat{Q}(\Omega -2cY)$ is
also BRST-trivial.

To prove the converse, i.e. that any state in the cohomology of
$\widehat Q$
maps to a state in the cohomology of $Q'$, suppose that the state
$\widehat
V$ is in the cohomology of $\widehat Q$. By choosing the $b$ ghost
to annihilate the vacuum, one can write $\widehat V = V+ cW$
for some $V$ and $W$.
Then $\widehat Q V =
\frac{1}{2}\widehat{\lambda}_{\dot a}\widehat{\lambda}_{\dot
a} W$
implies $Q' V=0$ in the reduced Hilbert space. And
$\widehat V=\widehat Q \Lambda$
where $\Lambda =\Omega + c Y$
implies that $V= \widehat Q \Omega
-\frac{1}{2}\widehat{\lambda}_{\dot a}\widehat{\lambda}_{\dot
a} Y$, i.e.
$V= Q' \Omega$ in the reduced Hilbert space. So the
cohomology
of $Q'$ is equivalent to the cohomology of $\widehat Q$.

Now, it will be shown that the cohomology of
$Q^{'} = \widehat{\lambda}_{a}\widehat{d}_{a} +
{\lambda}_{\dot{a}}\widehat{ d}_{\dot{a}}$ is
equivalent to the cohomology
of $Q=\lambda^{\alpha}d_{\alpha}$ in a Hilbert space
independent of $S_a$,
where $\lambda^{\alpha}$ is a pure spinor. To do this,
it is convenient to define an antichiral spinor ${r}_{\dot a}$ which
satisfies ${r}_{\dot a}{\lambda}_{\dot a}=1$ and
$r_{\dot a} r_{\dot a}=0$. 
One can then use ${r}_{\dot a}$ to split
the fields $S_a$ and $\widehat{\lambda}_{a}$ as\footnote
{In order that the cohomology remain non-trivial after including
$r_{\dot a}$, states will only be allowed to depend on
$r_{\dot a}$ in the combination 
$(\sigma^j r)_a (\sigma^j \lambda)_b$.
If states could depend
arbitrarily on $r_{\dot a}$, the cohomology would become trivial since
$QV=0$ implies that $Q(\theta^{\dot a} r_{\dot a} V)=V.$}

\begin{eqnarray}
S_a &=& S^{1}_{a} + S^{2}_{a} \:,
\nonumber\\
\widehat{\lambda}_{a} &=& \widehat{\lambda}^{1}_{a} +
\widehat{\lambda}^{2}_{a} \:, 
\end{eqnarray}

\noindent where

\begin{eqnarray}
S^{1}_{a} = \frac{1}{2} (\sigma^{j}{\lambda})_{a}
(S\sigma^{j}{r}) \:, \:\:
S^{2}_{a} =\frac{1}{2}  (\sigma^{j}{r})_{a}
(S\sigma^{j}{\lambda}) \:,\nonumber\\
\widehat{\lambda}^{1}_{a} =\frac{1}{2}  (\sigma^{j}{\lambda})_{a}
(\widehat{\lambda} \sigma^{j}{r}) \:, \:\:
\widehat{\lambda}^{2}_{a} =\frac{1}{2}  (\sigma^{j}{r})_{a}
(\widehat{\lambda}
\sigma^{j}{\lambda}) \:.
\end{eqnarray}

\noindent The new fields have the anticommutation relations

\begin{eqnarray}
\{S^{1}_{a}, S^{2}_{b}\} &=& \frac{1}{2} (\sigma^{i}
{\lambda})_{a}(\sigma^{i}{r})_{b} \:,\nonumber\\
\{S^{1}_{a}, S^{1}_{b}\} &=& \{S^{2}_{a}, S^{2}_{b}\}
= 0\:.
\end{eqnarray}

\noindent And the charge $Q^{'}$, written in terms of
the new fields, reads

\begin{equation}
Q^{'} = \widehat{\lambda}^{1}_{a}d_{a} +
\widehat{\lambda}^{2}_{a}d_{a} +
2\widehat{\lambda}^{1}_{a}S^{2}_{a}\sqrt{P^+} +
2\widehat{\lambda}^{2}_{a}S^{1}_{a}\sqrt{P^+}
+ {\lambda}_{\dot{a}}{d}_{\dot{a}} -
\frac{2(S^{2}\sigma^{i}{\lambda})P^{i}}{\sqrt{P^+}} \:.
\end{equation}

Performing the similarity transformation

\begin{eqnarray}
Q^{'} \longrightarrow
e^{-\frac{d_{a}S^{2}_{a}}{2\sqrt{P^{+}}}}
Q^{'}
e^{+\frac{d_{a}S^{2}_{a}}{2\sqrt{P^{+}}}}
\nonumber\\
= {\lambda}_{\dot{a}}{d}_{\dot{a}} +
\widehat{\lambda}^{1}_{a}d_{a} +
2\widehat{\lambda}^{2}_{a}S^{1}_{a} {\sqrt{P^+}}
\:,
\end{eqnarray}

\noindent one obtains $Q^{'} =
\lambda^{\alpha}d_{\alpha} +
2\widehat{\lambda}^{2}_{a}S^{1}_{a}\sqrt{P^+}$,
where $\lambda^{\alpha}$
is a pure spinor defined by

\begin{equation}
[{\lambda}_{\dot{a}},
\lambda_{a} ] =
[{\lambda}_{\dot{a}},
\widehat{\lambda}^{1}_{a}] \:.
\end{equation}

\noindent Using the quartet argument,
the cohomology
of $Q^{'} = Q +
2\widehat{\lambda}^{2}_{a}S^{1}_{a}\sqrt{P^+}$ is
equivalent to the cohomology of
$Q=\lambda^{\alpha}d_{\alpha}$ in the
Hilbert space independent of
$\widehat{\lambda}^{2}_{a}$ and $S^{1}_{a}$, and independent of
their conjugate momenta $\widehat{w}^{1}_{a}$ and
$S^{2}_{a}$. Therefore, the  action and BRST operator for the
Brink-Schwarz superparticle in semi-light-cone gauge
are equivalent to the
action

\begin{equation}
S=\int d\tau \left(\dot{x}^{m}P_{m}
-\frac{1}{2}P^{m}P_{m}
+
i\dot{\theta}^{\alpha}p_{\alpha} +
i\dot{\lambda}^{\alpha}w_{\alpha}\right)\:,
\end{equation}

\noindent and the BRST operator $Q=\lambda^{\alpha}d_{\alpha}$
where
$(\lambda\gamma^{m}\lambda)=0$.

\section{ Equivalence Proof for Green-Schwarz superstring}

\subsection{Green-Schwarz superstring in the
semi-light-cone gauge}

As in the covariant Brink-Schwarz superparticle
action, the GS superstring
action contains first-class and second-class
constraints which are difficult
to separate in a covariant manner. Quantization can be
performed
in a conformally invariant manner by gauge-fixing
$\kappa$-symmetry using
the condition
$(\gamma^{+}\theta)_{\alpha}=0$, assuming that
$\partial X^+\neq 0$.
In this semi-light-cone gauge, the GS action is
written as

\begin{equation}
S = \frac{1}{\pi} \int d^{2}z
\left[\frac{1}{2}\partial X^{m}
\overline{\partial} X_{m} +
\frac{1}{2}S_{a} \overline{\partial} S_{a} +
\mbox{anti-holomorphic terms}\right] \:,
\label{semi-gs}
\end{equation}

\noindent where $S_a = \sqrt{\frac{\partial
X^{+}}{2}}(\gamma^{-}\theta)_{a}$
is a chiral $SO(8)$ spinor. The anti-holomorphic terms
in (\ref{semi-gs})
depend if one is discussing the Type II or heterotic
superstring,
and will be ignored in this paper.

In semi-light-cone gauge, one can construct
a BRST charge in the
standard manner
as
\begin{equation}
Q= \int dz (c T_m + b c \partial c)
\end{equation}
with the action
\begin{equation}
S = \frac{1}{\pi} \int d^{2}z
\left[\frac{1}{2}\partial X^{m}
\overline{\partial} X_{m} +
\frac{1}{2}S_{a} \overline{\partial} S_{a} +
b\overline{\partial}c +
\mbox{anti-holomorphic terms}\right] \:,
\label{acao-gs}
\end{equation}
where

\begin{equation}
T_m = -\partial X^{-}\partial
X^{+} +
\partial X^{i}\partial X^{i}
-\frac{1}{2}S_{a}\partial S_{a}
+ \frac{1}{2}\partial^{2}(\log\partial X^{+}) \:
\end{equation}
is the stress tensor. The term
$\frac{1}{2}\partial^{2}(\log\partial X^{+})$ in the
stress tensor
comes from the non-covariant gauge-fixing and, as
will
be shown below, is necessary both for quantum
conformal
invariance and quantum Lorentz invariance.\footnote{Although there
are many papers which discuss anomalies in semi-light-cone-gauge
for the GS superstring \cite{semi-light}, we are not aware of any 
discussion
which uses this BRST method.}
Using the OPE's

\begin{eqnarray}
X^{m}(y,\overline{y}) X^{n}(z,\overline{z}) &\longrightarrow&
\frac{1}{2}\eta^{mn}\log\mid y-z\mid^{2} \:,
\nonumber\\
S_{a}(y)S_{b}(z) &\longrightarrow&
\frac{\delta_{ab}}{y-z} \:,
\label{ope}
\end{eqnarray}

\noindent one finds that $T_m$ has central charge
$c=26$, so $Q$ is nilpotent at the quantum level.

Although Lorentz invariance is not manifest, one can construct
Lorentz
generators that commute with $Q$. The holomorphic components
of the currents for these generators are

\begin{eqnarray}
N^{ij} &=& -X^{i}\partial X^{j} +X^{j}\partial X^{i}
-\frac{1}{4}(S\sigma^{ij}S) \:,
\nonumber \\
N^{+-} &=& -\frac{1}{2}X^{+}\partial X^{-}
+\frac{1}{2}X^{-}\partial X^{+}  \:,\nonumber \\
N^{i+} &=& -X^{i}\partial X^{+} + X^{+}\partial
X^{i} \:,\nonumber\\
N^{i-} &=& - X^{i}\partial X^{-} + X^{-}\partial
X^{i} -
\frac{(S\sigma^{i})_{\dot{a}}(S\sigma^{j})_{\dot{a}}\partial
X^{j}}{2\partial X^{+}} \:.
\end{eqnarray}

\noindent As in
the superparticle case,
the algebra closes up to a BRST-trivial operator:

\begin{equation}
\left[\int dy N^{i-}(y),\int dz N^{j-}(z)\right] =
\left[ Q,
\int dz \left[ -
\frac{b(S\sigma^{i})_{\dot{a}}(S\sigma^{j})_{\dot{a}}}{(\partial
X^{+})^2}\right](z)
\right]\:.
\end{equation}

So after including the term
$\frac{1}{2}\partial^{2}(\log\partial X^{+})$ in
$T_m$,
the Lorentz algebra closes on on-shell states up to a
gauge transformation.

\subsection{Equivalence between Green-Schwarz
and pure spinor
formalisms}

As with the superparticle, we shall add a new
pair of
fermionic degrees of
freedom $(\theta^{\alpha},p_{\alpha})$ not related to
$S_a$, and a new gauge invariance which allows
$\theta^\alpha$
to be gauged to zero. The new gauge invariance will be generated by
sixteen
first-class constraints $\widehat d_\alpha$ constructed from the $S_a$
variables and the first and second-class GS
constraints $d_\alpha$.
For the GS superstring, the holomorphic first and
second-class constraints are

\begin{equation}
d_{\alpha} = p_{\alpha} +
(\theta\gamma^{m})_{\alpha}(2\partial X_{m} +
(\theta\gamma_{m}{\partial}\theta)) \:,
\end{equation}
which satisfy the OPE's
\begin{equation}
d_{\alpha}(y) d_{\beta}(z) \longrightarrow
\frac{4\gamma^{m}_{\alpha \beta}
\Pi_{m}}{y-z}\:
\end{equation}
where
$\Pi^{m} = \partial X^{m} -
(\theta\gamma^{m}\partial \theta)$.

By combining these constraints with
the $S_a$ variables satisfying the OPE's
\begin{equation}
S_{a}(y)S_{b}(z) \longrightarrow
\frac{\delta_{ab}}{y-z} \:,
\end{equation}
one can construct the sixteen first-class constraints

\begin{eqnarray}
\widehat{d}_{a} &=& d_{a} + 2S_{a} \sqrt{\Pi^+}
\:,\nonumber\\
\widehat{ d}_{\dot{a}} &=& {d}_{\dot{a}} -
\frac{2\Pi^{i}(S\sigma^{i})_{\dot{a}}}{\sqrt{\Pi^{+}}}
+
\frac{(S\sigma^{jk}S)(\overline{\sigma}^{jk}\partial
{\theta})_{\dot{a}}}{4\Pi^{+}}
- \frac{4\partial^{2}{\theta}_{\dot{a}}}{\Pi^+} +
\frac{2\partial \Pi^{+} \partial
{\theta}_{\dot{a}}}{(\Pi^+)^2}\:,
\label{hat-gs}
\end{eqnarray}

\noindent where $a, \dot{a} = 1$ to 8.

\noindent The constraints of (\ref{hat-gs}) obey the
first-class algebra:

\begin{eqnarray}
\widehat{ d}_{\dot{a}}(y) \widehat{ d}_{\dot{b}}(z)
\longrightarrow
\frac{\widetilde{T}(z)\delta_{\dot{a}\dot{b}}}{y-z} \:,
\nonumber\\
\widehat{ d}_{\dot{a}}(y) \widehat{d}_{b}(z) \longrightarrow
\mbox{regular} \:,
\nonumber\\
\widehat{d}_{a}(y) \widehat{d}_{b}(z) \longrightarrow
\mbox{regular}\:,
\end{eqnarray}

\noindent where

\begin{eqnarray}
\widetilde{T} = -4\Pi^{-} +\frac{8S_{a}\partial
\theta_{a}}{\sqrt{\Pi^{+}}} -
\frac{8\Pi^{i}(S\sigma^{i}\partial
{\theta})}{(\Pi^{+})^{3/2}} -
\frac{2S_{a}\partial S_{a}}{\Pi^{+}} +
\frac{4\Pi^{i}\Pi^{i}}{\Pi^{+}}
\nonumber\\
+\frac{4(S\sigma^{i}\partial
{\theta})(S\sigma^{i}\partial{\theta})}
{(\Pi^{+})^{2}} -
\frac{16\partial^{2}{\theta}_{\dot{c}}\partial
{\theta}_{\dot{c}}}{(\Pi^{+})^{2}}
- \frac{2\partial^{2} (\log \Pi^{+})}{\Pi^{+}} \:.
\end{eqnarray}

\noindent $\widetilde{T}$ is also a first-class quantity
which satisfies
the OPE's

\begin{eqnarray}
\widetilde{T}(y)\widetilde{T}(z) &\longrightarrow&
\mbox{regular} \:,\nonumber\\
\widetilde{T}(y) \: \widehat{ d}_{\dot{a}}(z) &\longrightarrow&
\mbox{regular}\:,
\nonumber\\
\widetilde{T}(y) \: \widehat{d}_{a}(z) &\longrightarrow&
\mbox{regular}\:.
\end{eqnarray}

\noindent The following OPE's were used in these
calculations:

\begin{eqnarray}
d_{a}(y) d_{b}(z) \longrightarrow -\frac{4\delta_{ab}
\Pi^{+}}{y-z} \:,
\nonumber\\
d_{a}(y) {d}_{\dot{a}}(z) \longrightarrow
\frac{4\sigma^{i}_{a\dot{a}}
\Pi^{i}}{y-z} \:,\nonumber\\
{d}_{\dot{a}}(y) {d}_{\dot{b}}(z) \longrightarrow -
\frac{4\delta_{\dot{a}\dot{b}}
\Pi^{-}}{y-z}\:,\nonumber\\
d_{a}(y) \Pi^{+}(z) \longrightarrow \mbox{regular}
\:,\nonumber\\
d_{a}(y) \Pi^{-}(z) \longrightarrow \frac{4 \partial
\theta_{a}}{y-z} \:,
\nonumber\\
d_{a}(y) \Pi^{i}(z) \longrightarrow 
\frac{2 (\sigma^{i}\partial {\theta})_{a}}{y-z}\:,
\nonumber\\
{d}_{\dot{a}}(y) \Pi^{+}(z) \longrightarrow 
\frac{4 \partial {\theta}_{\dot{a}}}{y-z}
\:,\nonumber\\
{d}_{\dot{a}}(y) \Pi^{-}(z) \longrightarrow
\mbox{regular}\:, \nonumber\\
{d}_{\dot{a}}(y) \Pi^{i}(z) \longrightarrow 
\frac{2 (\sigma^{i}\partial
\theta)_{\dot{a}}}{y-z}\:.
\label{ope1}
\end{eqnarray}

Using the first-class constraints $\widehat d_\alpha$ and
$\widetilde T$,
one can construct
a nilpotent BRST operator in the usual manner as
\begin{equation}
\widehat{Q} =\int dz( c\widetilde{T} + \widehat{\lambda}_{\dot{a}}
\widehat{ d}_{\dot{a}} +
\widehat{\lambda}_{a} \widehat{d}_{a} -
\frac{1}{2}\widehat{\lambda}_{\dot{a}}\widehat{\lambda}_{\dot{a}}b)
\:,
\label{qhat-gs}
\end{equation}

\noindent with the worldsheet action

\begin{equation}
S = \frac{1}{\pi} \int d^{2}z
\left[\frac{1}{2}\partial X^{m}
\overline{\partial} X_{m} +
\frac{1}{2}S_{a} \overline{\partial} S_{a} +
p_{\alpha}\overline{\partial}
\theta^{\alpha} +
\widehat{w}_{\alpha}
\overline{\partial}\widehat{\lambda}^{\alpha}
+ b\overline{\partial}{c}\right] \:,
\end{equation}

\noindent where $(b,c)$ are the fermionic ghosts
for the
bosonic constraint $\widetilde{T}$,
and $(\widehat{\lambda}^{\alpha},\widehat w_\alpha)$ are
unconstrained
bosonic
spinorial ghosts.

As in the superparticle, one could rescale
$b\rightarrow \frac{b}{\Pi^+}$ and $c\rightarrow
c\Pi^+$ so that $c$
multiplies the standard stress tensor in the BRST
operator.
This can be done at the quantum level using the
similarity
transformation
\begin{equation}
Q \rightarrow e^{-\int dz bc\log(\Pi^+)} Q
e^{+\int dz bc\log(\Pi^+)}  =4\int dz (c T_m + b
c\partial c + \ldots)
\end{equation}
where $\int dz (c T_m + b c \partial c)$ is the BRST
operator of (\ref{acao-gs})
and $\ldots$ involves the new variables
$(\theta^\alpha,p_\alpha)$ and
$(\widehat{\lambda}^\alpha, \widehat{w}_\alpha)$.
However, it will be more convenient to not rescale the
$(b,c)$ ghosts
so that $\widehat Q$ has the simple structure of
(\ref{qhat-gs}). The usual stress tensor can be
obtained from $\widehat Q$ by

\begin{eqnarray}
T&=&\{\widehat{Q}, \frac{1}{4}
b\Pi^{+} -\widehat{w}_{\alpha}\partial
\theta^{\alpha} \} \nonumber\\
&=& \Pi^{m}\Pi_{m}  -\frac{1}{2} S_{a}\partial S_{a} -
d_{\alpha}\partial \theta^{\alpha}
- \widehat{w}_{\alpha}\partial\widehat{\lambda}^{\alpha}
 - b\partial c -
\frac{1}{2}\partial^{2}(\log \Pi^{+})\:,
\label{t}
\end{eqnarray}

\noindent so $\frac{1}{4}b\Pi^{+} -\widehat w_\alpha \partial
\theta^\alpha$ plays the
role of the usual $b$ ghost. Using the OPE's of
(\ref{ope}) and (\ref{ope1}), one can verify
that $T$ has no central charge. Note that the $(b,c)$
ghosts in (\ref{t}) have
not been rescaled so they carry conformal weight
$(1,0)$ instead
of $(2,-1)$.

As before, Lorentz invariance is not manifest but
Lorentz
generators can be constructed which leave $\widehat Q$
invariant.
The holomorphic components of the currents for these generators are

\begin{eqnarray}
N^{ij} &=& -X^{i}\partial X^{j} + X^{j}\partial
X^{i} + \frac{1}{2}
(\widehat{\lambda}
\sigma^{ij} \widehat{w}) + \frac{1}{2} (\widehat{\lambda}
\overline{\sigma}^{ij} \widehat{ w})
+ \frac{1}{2} (\theta \sigma^{ij} p) + \frac{1}{2}
({\theta}\overline{\sigma}^{ij}{p}) - \frac{1}{4} (S
\sigma^{ij} S) \:,\nonumber \\
N^{+-} &=& -\frac{1}{2}X^{+}\partial X^{-} +
\frac{1}{2}X^{-}\partial X^{+} + \frac{1}{2}
\widehat{\lambda}_{a} \widehat{w}_{a} - \frac{1}{2}
\widehat{\lambda}_{\dot{a}}
\widehat{ w}_{\dot{a}}
+ \frac{1}{2} \theta_{a} p_{a} - \frac{1}{2}
{\theta}_{\dot{a}}{p}_{\dot{a}} -
bc \:,\nonumber \\
N^{i+} &=& -X^{i}\partial X^{+} + X^{+}\partial
X^{i} + (\widehat{w}
\sigma^{i} \widehat{\lambda}) -
(p \sigma^{i}{\theta}) \:,\nonumber\\
N^{i-} &=& -X^{i}\partial X^{-} + X^{-}\partial
X^{i} -
\frac{(S\sigma^{i})_{\dot{a}}(S\sigma^{j})_{\dot{a}}\Pi^{j}}{2\Pi^{+}}
+
\frac{(S\sigma^{i})_{\dot{a}}(S\sigma^{j})_{\dot{a}}
(S\sigma^{j}\partial{\theta})}{3(\Pi^{+})^{3/2}} +
\frac{2(\partial
S\sigma^{i}\partial{\theta})}{(\Pi^{+})^{3/2}}
\nonumber\\
&-& \frac{2(S\sigma^{i}\partial{\theta})\partial
\Pi^{+}}{(\Pi^{+})^{5/2}} +
\frac{b(S\sigma^{i}\widehat{\lambda})}{2\sqrt{\Pi^+}} +
\frac{2bc(S\sigma^{i}\partial
{\theta})}{(\Pi^{+})^{3/2}} -
\frac{2bc\Pi^{i}}{\Pi^{+}} \nonumber\\
&+& (\widehat{\lambda}\sigma^{i}\widehat{ w}) +
(\theta\sigma^{i}{p})\:.
\end{eqnarray}

\noindent Once again, they obey the usual Lorentz
algebra except for $N^{i-}$
with $N^{j-}$, which satisfies

\begin{equation}
\left[\int dy N^{i-}(y),\int dz N^{j-}(z)\right] =
\left[ \widehat{Q},
\int dz
\left[-\frac{b(S\sigma^{i})_{\dot{a}}(S\sigma^{j})_{\dot{a}}}
{4\Pi^{+}}\right](z)\right]\:.
\end{equation}
So the Lorentz algebra closes on on-shell states up to
a gauge transformation.

The BRST operator $\widehat Q$ will now be related to the pure
spinor BRST operator $Q=\int dz \lambda^\alpha d_\alpha$.
The first step is to relate the cohomology of
the BRST charge (\ref{qhat-gs})
to the cohomology of a charge $Q^{'} =
\int dz (\widehat{\lambda}_{a}\widehat{d}_{a} +
{\lambda}_{\dot{a}}\widehat{ d}_{\dot{a}})$ where
${\lambda}_{\dot{a}}$ has to satisfy
${\lambda}_{\dot{a}}{\lambda}_{\dot{a}}=0$.

Suppose a state
$V$ is in the cohomology of $Q'$,
which implies that $V$ is annihilated by
\begin{equation}
{Q''} =\int dz( \widehat{\lambda}_{a} \widehat d_{a} +
\widehat{\lambda}_{\dot{a}}\widehat{d}_{\dot{a}} )
\label{carga6}
\end{equation}
up to terms
proportional to
$\widehat{\lambda}_{\dot{a}}\widehat{\lambda}_{\dot{a}}$ or
its derivatives. So

\begin{equation}
Q^{''}V = \sum_{n=0}^{\infty} \partial^n
(\widehat{\lambda}_{\dot{a}}\widehat{\lambda}_{\dot{a}})
W_{(n)} \:,
\end{equation}

\noindent for some $W_{(n)}$ for $n=0$ to $\infty$.
In addition, suppose that $V$ has no poles
with ${\lambda}_{\dot a}{\lambda}_{\dot a}$, i.e. $V$ only
depends on ${w}_{\dot a}$ in combinations which commute
with
the constraint on ${\lambda}_{\dot a}$. Then, from the
relation

\begin{equation}
(Q^{''})^{2} = \int dy
\frac{(\widehat{\lambda}_{\dot{a}}\widehat{\lambda}_{\dot{a}})\widetilde
T}{2}
\:,
\end{equation}

\noindent it follows that

\begin{equation}
Q^{''}W_{(n)} = \int dy\frac{1}{2}\widetilde T(y)V(z)
\frac{(y-z)^n}{n!} \:.
\end{equation}

\noindent Using the above equations,
it is easy to
check that the state $\widehat{V} = V + 2\sum_n(\partial^n
c)W_{(n)}$ is
annihilated by $\widehat{Q}$. Also,
if a state $V$ is BRST trivial up to terms
$\partial^n(\widehat{\lambda}_{\dot{a}}\widehat{\lambda}_{\dot{a}})$,
i.e.

\begin{equation}
V = Q^{''}\Omega +
\sum_{n=0}^\infty
\partial^n
(\widehat{\lambda}_{\dot{a}}\widehat{\lambda}_{\dot{a}})Y_{(n)}
\:,
\end{equation}

\noindent for some $Y_{(n)}$, then $\widehat{V}$ is also
BRST
trivial with respect to
$\widehat{Q}$ since

\begin{equation}
\widehat{V} = V + 2\sum_n(\partial^n c)W_{(n)} =
\widehat{Q}(\Omega -
2\sum_n (\partial^n c) Y_{(n)})\:.
\end{equation}

To complete the proof, one needs to show that any state in the
cohomology of $\widehat Q$ can be mapped to a state in the cohomology
of $Q'$. To show this, first note that
any ghost-number one state $\widehat V$ in the cohomology
of $\widehat Q$
with non-zero $P^+$ momentum can be expressed as $\widehat
V = V+cW$ for some $V$ and $W$ which are independent of the
$(b,c)$ ghosts.
This is because in light-cone gauge, the constraints
$\widetilde T$ and
$\widehat d_\alpha$
can be used to gauge away dependence on all variables except for
$X^j$, $S^a$ and
the zero mode of $X^+$
in the integrated light-cone vertex operator
$\int dz V_{LC}(X^j,S^a, X^+_0)$. So using the standard
DDF construction,
one can define a BRST-invariant vertex operator
$\int dz V_{DDF}(X^j,S^a,X^+, \theta^\alpha)$
such that
$V_{DDF}$ coincides with $V_{LC}$ when $\partial X^+ =
\theta^\alpha=0$.
Since $\int dz V_{DDF}$  is BRST-invariant,
$\widehat Q V_{DDF} = \partial \widehat V$
for some $\widehat V$. From the structure of the BRST operator
$\widehat Q$ of (\ref{qhat-gs}), one learns that
$\widehat V= \widehat\lambda^\alpha V_\alpha + c W$
where $V_\alpha$ and $W$ are the double poles of $\widehat
d_\alpha$ and $\widetilde T$
with $V_{DDF}$. Since $ \partial (\widehat Q \widehat V)=
\widehat Q \widehat Q V_{DDF}=0$ and since there are
no constant worldsheet fields,
$\widehat Q\widehat V=0$. Therefore,
$\widehat V= \widehat\lambda^\alpha V_\alpha + cW$ is a
ghost-number one vertex
operator
in the BRST cohomology which represents the light-cone
state
$\int dz V_{LC}$.

Since $\widehat Q( V + cW)=0$ implies that
$Q'' V = \frac{1}{2}\widehat{\lambda}_{\dot
a}\widehat{\lambda}_{\dot a} W$,
$Q' V=0$ in the reduced Hilbert space. And
$\widehat V=\widehat Q \Lambda$
where $\Lambda =\Omega + c Y$
implies that $V= Q'' \Omega
-\frac{1}{2}\widehat{\lambda}_{\dot a}\widehat{\lambda}_{\dot
a} Y$, i.e.
$V= Q' \Omega$ in the reduced Hilbert space. So the
cohomology
of $Q'$ is equivalent to the cohomology of $\widehat Q$
for states
with non-zero $P^+$ momentum.\footnote{The equivalence
proof does
not hold for states of zero momentum since such states
cannot be described by light-cone vertex operators. For example, the
stress
tensor
$T=\{\widehat{Q},\frac{1}{4} b\Pi^{+} -\widehat{w}_{\alpha}\partial
\theta^{\alpha}
\}$ of
(\ref{t}) is
BRST trivial, but the stress tensor in the pure spinor
formalism
is in the BRST cohomology since there are no
states of negative ghost-number. }

To complete the equivalence proof, the relation between
$Q^{'}$ and the pure
spinor BRST operator $Q=\int dz \lambda^{\alpha}d_{\alpha}$
with
$(\lambda\gamma^{m}\lambda)=0$ has to be shown. For
this purpose, it is convenient to define an antichiral
spinor ${r}_{\dot a}$ satisfying ${r}_{\dot a}{r}_{\dot a}=0$ and
${r}_{\dot a}{\lambda}_{\dot a}=1$, and
to split the fields
$S_a$ and $\widehat{\lambda}_{a}$ into\footnote
{As in the superparticle, $r_{\dot a}$ will only be allowed
to appear in the combination $(\sigma^j r)_a (\sigma^j \lambda)_b$
so that the cohomology remains non-trivial.}

\begin{eqnarray}
S_a &=& S^{1}_{a} + S^{2}_{a} \:,
\nonumber\\
\widehat{\lambda}_{a} &=& \widehat{\lambda}^{1}_{a} +
\widehat{\lambda}^{2}_{a} \:, 
\end{eqnarray}

\noindent where

\begin{eqnarray}
S^{1}_{a} = \frac{1}{2}(\sigma^{j}{\lambda})_{a}
(S\sigma^{j}{r}) \:, \:\:
S^{2}_{a} = \frac{1}{2}(\sigma^{j}{r})_{a}
(S\sigma^{j}{\lambda}) \:,\nonumber\\
\widehat{\lambda}^{1}_{a} = \frac{1}{2} (\sigma^{j}{\lambda})_{a}
(\widehat{\lambda} \sigma^{j}{r}) \:, \:\:
\widehat{\lambda}^{2}_{a} = \frac{1}{2} (\sigma^{j}{r})_{a}
(\widehat{\lambda}
\sigma^{j}{\lambda}) \:,
\end{eqnarray}

\noindent which have the OPE's:

\begin{eqnarray}
S^{1}_{a}(y)S^{2}_{b}(z) &\longrightarrow&
\frac{(\sigma^{i}
{\lambda})_{a}(\sigma^{i}{r})_{b}}{2(y-z)} \:,\nonumber\\
S^{1}_{a}(y)S^{1}_{b}(z) &=& S^{2}_{a}(y)S^{2}_{b}(z)
\longrightarrow
\mbox{regular}\:.
\end{eqnarray}

\noindent In terms of these fields,

\begin{eqnarray}
Q' &=& \int dz ({\lambda}_{\dot{a}} \widehat{ d}_{\dot{a}} +
\widehat{\lambda}_{a} \widehat{d}_{a} )
\nonumber\\
&=& \int dz({\lambda}_{\dot{a}} {d}_{\dot{a}} -\frac{2\Pi^{i}
(S^{2} \sigma^{i}
{\lambda})}{\sqrt{\Pi^+}} - 2\frac{:S^{1}_{a}
S^{2}_{a}:
{\lambda}_{\dot{a}} \partial {\theta}_{\dot{a}}}{
\Pi^+} +
\frac{(S^{2}\sigma^{i}{\lambda})(S^{2}\sigma^{i}\partial
{\theta})}{\Pi^+}
\nonumber\\
&-&
4\frac{\partial^{2}{\theta}_{\dot{a}}{\lambda}_{\dot{a}}}{\Pi^+}
+
2\frac{\partial \Pi^{+} \partial {\theta}_{\dot{a}}
{\lambda}_{\dot{a}}}{(\Pi^{+})^2} +
\widehat{\lambda}^{1}_{a} d_{a}
+  \widehat{\lambda}^{2}_{a}d_{a} \nonumber\\
&+& 2 \widehat{\lambda}^{1}_{a} S^{2}_{a}
\sqrt{\Pi^+} + 2
\widehat{\lambda}^{2}_{a} S^{1}_{a} \sqrt{\Pi^+}) \;,
\end{eqnarray}

\noindent where $::$ denotes normal ordering, i.e.
$:S_a^1 S_b^2 (z):\equiv
\int dy S_a^1(y) S_b^2(z) (y-z)^{-1}$.
Performing the similarity
transformation
$Q^{'} \rightarrow e^{-\int dz A}Q^{'}e^{\int dz A} $
where $ A= \frac{d_{a} S^{2}_{a}}{2\sqrt{\Pi^{+}}}$,
one obtains

\begin{eqnarray}
Q^{'} &=& \int dz ( {\lambda}_{\dot{a}} {d}_{\dot{a}} -
\frac{2:S^{1}_{a} S^{2}_{a}:
{\lambda}_{\dot{a}} \partial {\theta}_{\dot{a}}}{
\Pi^+} +
\frac{4(\partial{\theta}_{\dot{a}}{\lambda}_{\dot{a}})
(\partial{\lambda}_{\dot{b}}{r}_{\dot{b}})}{\Pi^+} -
\frac{2\partial \Pi^{+} \partial {\theta}_{\dot{a}}
{\lambda}_{\dot{a}}}{(\Pi^{+})^2} \nonumber\\
&+&  \widehat{\lambda}^{1}_{a} d_{a} +
2 \widehat{\lambda}^{2}_{a} S^{1}_{a} \sqrt{\Pi^+} )\:.
\end{eqnarray}

\noindent
To simplify further
the BRST charge, one performs the additional
similarity transformation
$Q^{'} \rightarrow e^{-\int dz B}Q^{'}e^{\int dz B} $
where
\begin{equation}
B= -\frac{\partial\Pi^{+}}{2\Pi^{+}} +
\frac{1}{2}:S^{1}_{a}S^{2}_{a}:
\log{\Pi^{+}}
+\frac{4(\partial{\theta}_{\dot{a}}{\lambda}_{\dot{a}})
(\partial{\theta}_{\dot{b}}{r}_{\dot{b}})}{\Pi^+}\:,
\end{equation}
\noindent to obtain

\begin{equation}
Q^{'} = \int dz ({\lambda}_{\dot{a}} {d}_{\dot{a}}
+ \widehat{\lambda}^{1}_{a}
d_{a} + 2 \widehat{\lambda}^{2}_{a} S^{1}_{a}) \:.
\end{equation}

\noindent So after performing these
similarity transformations, $ Q^{'} =\int dz(
\lambda^{\alpha}d_{\alpha} + 2
\widehat{\lambda}^{2}_{a} S^{1}_{a})$ where $\lambda^{\alpha}$
is a pure spinor defined
by

\begin{equation}
[{\lambda}_{\dot{a}}, \lambda_{a}]
=
[{\lambda}_{\dot{a}},
\widehat{\lambda}^{1}_{a}]
\:.
\end{equation}

\noindent Making use of the standard quartet argument,
the cohomology of
$Q^{'} = Q + 2\int dz  \widehat{\lambda}^{2}_{a} S^{1}_{a}$
is equivalent to the
cohomology of $Q=\int dz \lambda^{\alpha}d_{\alpha}$ in a
Hilbert space independent
of $S^{1}_{a}$, $\widehat{\lambda}^{2}_{a}$ and their conjugate
momenta $S^{2}_{a}$
and $\widehat{w}^{1}_{a}$. So, it has been shown that the
Green-Schwarz superstring action  and BRST operator
in semi-light-cone gauge are equivalent to the action

\begin{equation}
S = \frac{1}{\pi}\int d^{2}z \left[\frac{1}{2}\partial
X^{m}\overline{\partial}X_{m} + p_{\alpha}
\overline{\partial}\theta^{\alpha}
+w_{\alpha} \overline{\partial}\lambda^{\alpha}
+
\mbox{anti-holomorphic terms}\right]\:,
\label{action-ps}
\end{equation}

\noindent and BRST operator $Q=\int dz \lambda^{\alpha}d_{\alpha}$
where $\lambda^{\alpha}$
is a pure spinor.

\acknowledgments

We would like to thank Daniel Nedel and Sasha Polyakov for useful
discussions.
N.B. would also like to thank CNPq grant 300256/94-9, Pronex 
66.2002/1998-9
and FAPESP grant 99/12763-0 for financial support, and D.Z.M. would 
like
to thank FAPESP grant 00/13845-9 for financial support.

\end{document}